\documentclass[amsmath,amssymb,twocolumn,prl,showpacs,nofootinbib,floatfix,superscriptaddress]{revtex4-1}
\usepackage{graphics}
\usepackage{dcolumn}
\usepackage{bm}
\usepackage{longtable}
\usepackage[pdftex]{graphicx}
\usepackage{dcolumn}
\usepackage{footnote}
\usepackage{amssymb}
\usepackage{multirow}
\usepackage{natbib}

\begin{document}
\title{Classical-Nova Contribution to the Milky Way's $^{26}$Al Abundance: Exit Channel of the Key $^{25}$Al($p,\gamma$)$^{26}$Si Resonance}
\author{M. B. Bennett}
\email{bennettm@nscl.msu.edu}
\affiliation{Department of Physics and Astronomy, Michigan State University, East Lansing, Michigan 48824, USA}
\affiliation{National Superconducting Cyclotron Laboratory, Michigan State University, East Lansing, Michigan 48824, USA}
\author{C. Wrede}
\email{wrede@nscl.msu.edu}
\affiliation{Department of Physics and Astronomy, Michigan State University, East Lansing, Michigan 48824, USA}
\affiliation{National Superconducting Cyclotron Laboratory, Michigan State University, East Lansing, Michigan 48824, USA}
\affiliation{Department of Physics, University of Washington, Seattle, Washington 98195, USA}
\author{K. A. Chipps}
\affiliation{Department of Physics, Colorado School of Mines, Golden, Colorado 08401, USA}
\author{J. Jos{\'e}}
\affiliation{Departament F{\'i}sica i Enginyeria Nuclear (UPC) and Institut d'Estudis Espacials de Catalunya (IEEC), E-08034 Barcelona, Spain}
\author{S. N. Liddick}
\affiliation{Department of Chemistry, Michigan State University, East Lansing, Michigan 48824, USA}
\affiliation{National Superconducting Cyclotron Laboratory, Michigan State University, East Lansing, Michigan 48824, USA}
\author{M. Santia}
\affiliation{Department of Physics and Astronomy, Michigan State University, East Lansing, Michigan 48824, USA}
\affiliation{National Superconducting Cyclotron Laboratory, Michigan State University, East Lansing, Michigan 48824, USA}
\author{A. Bowe}
\affiliation{Department of Physics and Astronomy, Michigan State University, East Lansing, Michigan 48824, USA}
\affiliation{National Superconducting Cyclotron Laboratory, Michigan State University, East Lansing, Michigan 48824, USA}
\affiliation{Physics Department, Kalamazoo College, Kalamazoo, Michigan 49006, USA}
\author{A. A. Chen}
\affiliation{Department of Physics and Astronomy, McMaster University, Hamilton, ON L8S 4M1, Canada}
\author{N. Cooper}
\affiliation{Department of Physics and Wright Nuclear Structure Laboratory, Yale University, New Haven, Connecticut 06520, USA}
\author{D. Irvine}
\affiliation{Department of Physics and Astronomy, McMaster University, Hamilton, ON L8S 4M1, Canada}
\author{E. McNeice}
\affiliation{Department of Physics and Astronomy, McMaster University, Hamilton, ON L8S 4M1, Canada}
\author{F. Montes}
\affiliation{National Superconducting Cyclotron Laboratory, Michigan State University, East Lansing, Michigan 48824, USA}
\affiliation{Joint Institute for Nuclear Astrophysics, Michigan State University, East Lansing, Michigan 48824, USA}
\author{F. Naqvi}
\affiliation{Department of Physics and Wright Nuclear Structure Laboratory, Yale University, New Haven, Connecticut 06520, USA}
\author{R. Ortez}
\affiliation{Department of Physics and Astronomy, Michigan State University, East Lansing, Michigan 48824, USA}
\affiliation{National Superconducting Cyclotron Laboratory, Michigan State University, East Lansing, Michigan 48824, USA}
\affiliation{Department of Physics, University of Washington, Seattle, Washington 98195, USA}
\author{S. D. Pain}
\affiliation{Oak Ridge National Laboratory, Oak Ridge, Tennessee 37831, USA}
\author{J. Pereira}
\affiliation{National Superconducting Cyclotron Laboratory, Michigan State University, East Lansing, Michigan 48824, USA}
\affiliation{Joint Institute for Nuclear Astrophysics, Michigan State University, East Lansing, Michigan 48824, USA}
\author{C. Prokop}
\affiliation{Department of Chemistry, Michigan State University, East Lansing, Michigan 48824, USA}
\affiliation{National Superconducting Cyclotron Laboratory, Michigan State University, East Lansing, Michigan 48824, USA}
\author{J. Quaglia}
\affiliation{Department of Electrical Engineering, Michigan State University, East Lansing, Michigan 48824, USA}
\affiliation{Joint Institute for Nuclear Astrophysics, Michigan State University, East Lansing, Michigan 48824, USA}
\affiliation{National Superconducting Cyclotron Laboratory, Michigan State University, East Lansing, Michigan 48824, USA}
\author{S. J. Quinn}
\affiliation{Department of Physics and Astronomy, Michigan State University, East Lansing, Michigan 48824, USA}
\affiliation{National Superconducting Cyclotron Laboratory, Michigan State University, East Lansing, Michigan 48824, USA}
\affiliation{Joint Institute for Nuclear Astrophysics, Michigan State University, East Lansing, Michigan 48824, USA}
\author{S. B. Schwartz}
\affiliation{Department of Physics and Astronomy, Michigan State University, East Lansing, Michigan 48824, USA}
\affiliation{National Superconducting Cyclotron Laboratory, Michigan State University, East Lansing, Michigan 48824, USA}
\affiliation{Geology and Physics Department, University of Southern Indiana, Evansville, Indiana 47712, USA}
\author{S. Shanab}
\affiliation{Department of Physics and Astronomy, Michigan State University, East Lansing, Michigan 48824, USA}
\affiliation{National Superconducting Cyclotron Laboratory, Michigan State University, East Lansing, Michigan 48824, USA}
\author{A. Simon}
\affiliation{National Superconducting Cyclotron Laboratory, Michigan State University, East Lansing, Michigan 48824, USA}
\affiliation{Joint Institute for Nuclear Astrophysics, Michigan State University, East Lansing, Michigan 48824, USA}
\author{A. Spyrou}
\affiliation{Department of Physics and Astronomy, Michigan State University, East Lansing, Michigan 48824, USA}
\affiliation{National Superconducting Cyclotron Laboratory, Michigan State University, East Lansing, Michigan 48824, USA}
\affiliation{Joint Institute for Nuclear Astrophysics, Michigan State University, East Lansing, Michigan 48824, USA}
\author{E. Thiagalingam}
\affiliation{Department of Physics and Astronomy, McMaster University, Hamilton, ON L8S 4M1, Canada}
%

\begin{abstract}

Classical novae are expected to contribute to the 1809-keV Galactic $\gamma$-ray emission by producing its precursor $^{26}$Al, but the yield depends on the thermonuclear rate of the unmeasured $^{25}$Al($p,\gamma$)$^{26}$Si reaction. Using the $\beta$ decay of $^{26}$P to populate the key $J^{\pi}=3^+$ resonance in this reaction, we report the first evidence for the observation of its exit channel via a $1741.6 \pm 0.6 (\textrm{stat}) \pm 0.3 (\textrm{syst})$ keV primary $\gamma$ ray, where the uncertainties are statistical and systematic, respectively. By combining the measured $\gamma$-ray energy and intensity with other experimental data on $^{26}$Si, we find the center-of-mass energy and strength of the resonance to be $E_r = 414.9 \pm 0.6(\textrm{stat}) \pm 0.3 (\textrm{syst}) \pm 0.6(\textrm{lit.})$ keV and $\omega\gamma = 23 \pm 6 (\textrm{stat})^{+11}_{-10}(\textrm{lit.})$ meV, respectively, where the last uncertainties are from adopted literature data. We use hydrodynamic nova simulations to model $^{26}$Al production showing that these measurements effectively eliminate the dominant experimental nuclear-physics uncertainty and we estimate that novae may contribute up to 30 \% of the Galactic $^{26}$Al.
\vskip\baselineskip

\noindent \hfill \scriptsize{PACS numbers: 23.20.Lv, 25.40.Lw, 26.30.Ca, 27.30.+t}

\end{abstract}

\maketitle

Gamma-ray telescopes pointed at the Milky Way have detected a diffuse 1809-keV line that is characteristic of $^{26}$Al decay ($\tau = 1.0$ Ma) \cite{ma84apj,sh85apj,ma87apj,di95aa,di06nat}. Observation of this line provides direct evidence for ongoing nucleosynthesis processes contributing to the interstellar medium and maintaining a total steady-state $^{26}$Al mass of $2.7 \pm 0.7$ $M_{\odot}$ \cite{di06nat,wa09aa}. The abundance of $^{26}$Al in protoplanetary disks orbiting young stars may influence the formation of habitable planetary systems such as our own because, in sufficient quantities, the energy released by its \textit{in situ} decay can heat planetesimals inducing differentiation and water sublimation \cite{ur55pnas,sr99sci,ti12conf}. The inhomogeneous spatial distribution of the 1809-keV emission across the Milky Way suggests that the outflows of massive stars and their supernovae are the primary sites for $^{26}$Al production \cite{pr96pr}. In the limit where secondary sites such as classical novae and asymptotic giant branch stars are well understood, one can use the $^{26}$Al line to estimate the rate of core-collapse supernovae in the Milky Way \cite{di06nat} or compare with the $^{60}$Fe gamma-ray line intensity \cite{wa07aa} to benchmark simulations of nucleosynthesis in models of massive-star evolution and death \cite{li06apj}.

Classical novae are thermonuclear explosions on hydrogen-accreting white-dwarf stars that have been estimated to contribute up to 0.4 $M_{\odot}$ to the Galactic $^{26}$Al inventory \cite{jo97apj}. This contribution needs to be quantified accurately for the intrinsic study of classical novae and because it could present a significant background to the massive-star component. Fortunately, modeling of nucleosynthesis in novae is relatively advanced and now includes experimental constraints on most of the essential thermonuclear reaction rates \cite{il10npaII}, which are primarily resonant radiative proton captures at peak temperatures between 0.1 and 0.4 GK. For example, the direct production mechanism for $^{26}$Al, the $^{25}$Mg($p,\gamma$)$^{26}$Al reaction, is well studied experimentally \cite{il90npa,il96prc,il10npaIII,st12plb} because the reactants are stable. A recent experiment using a $^{26}$Al rare isotope beam has reduced the uncertainty in the rate of the direct destruction mechanism, the $^{26}$Al($p,\gamma$)$^{27}$Si reaction \cite{ru06prl}. The dominant outstanding experimental nuclear-physics uncertainty lies in the thermonuclear rate of the $^{25}$Al($p,\gamma$)$^{26}$Si reaction ($Q = 5513.8 \pm 0.5$ keV \cite{pa05prc,er09prc,au12cpc}), which bypasses production of the $^{26}$Al ground state \cite{il96prc} and, therefore, reduces the intensity of 1809-keV gamma ray emission.

The $^{25}$Al($p,\gamma$)$^{26}$Si rate depends on the center-of-mass energies and strengths of $^{26}$Si resonances (for a recent summary see Ref. \cite{wr09prc}). Direct measurements of the resonance strengths using a $^{25}$Al ($\tau = 10.4$ s) beam are not yet possible at rare-isotope beam facilities due to insufficient intensities. Efforts to constrain the reaction rate indirectly by studying the proton-unbound states and mass of $^{26}$Si have included a variety of experimental nuclear-physics methods utilizing both stable and rare isotope beams \cite{ba02prc,pa04prc,th04epj,pa05prc,ba06prc,se07prc,kw08jkp,er09prc,pe09prc,ma10prc,ch10prc,de10pos,ch12prc,ko12pos}. In addition, reaction-rate evaluations employing available data and supplemented by shell-model calculations or information from the isospin mirror nucleus $^{26}$Mg have been conducted \cite{il96prc,il01apj,wr09prc,ri11prc,pa13prc}.

Currently, there are three known $^{26}$Si states (spin and parity 1$^+$, 0$^+$, and 3$^+$) that could potentially contribute to the $^{25}$Al($p,\gamma$)$^{26}$Si reaction rate as resonances at nova temperatures. The center-of-mass energy of the 1$^+$ resonance is $163.2 \pm 1.8$ keV based on the excitation energy derived from its gamma decay \cite{se07prc} and the $Q$ value. The 1$^+$ resonance likely only contributes to the total reaction rate at temperatures below 0.2 GK, where the $^{25}$Al($p,\gamma$)$^{26}$Si route bypassing $^{26}$Al is not strongly activated \cite{pa13prc}. The energy of the 0$^+$ resonance is not settled, but it appears to lie in the vicinity of 400 keV based on its population in the $^{24}$Mg($^3$He,$n$)$^{26}$Si reaction \cite{pa04prc,de10pos,ko12pos}. The strength of the 0$^+$ resonance is expected \cite{ba06prc} to be much lower than the nearby 3$^+$ resonance and, therefore, the 3$^+$ resonance likely dominates the reaction rate at the highest nova temperatures, where $^{25}$Al($p,\gamma$)$^{26}$Si is most active, making experimental constraints on the energy and strength of the 3$^+$ resonance essential. It has been argued \cite{wr09prc} that the 3$^+$ resonance energy is $412 \pm 2$ keV based on $^{26}$P beta-delayed proton decay \cite{th04epj} and other experimental data, corresponding to a $^{26}$Si excitation energy of $5926 \pm 2$ keV when combined with the $Q$ value; an excitation-energy value of $5927 \pm 4$ keV measured more recently using the $^{28}$Si($p,t$)$^{26}$Si$^*$($p$)$^{25}$Al reaction \cite{ch10prc} is in good agreement. The proton-decay partial width of the 3$^+$ resonance has been determined to be $\Gamma_p = 2.9 \pm 1.0$ eV using the $^{25}$Al($d,n$)$^{26}$Si$^*$($p$)$^{25}$Al reaction \cite{pe09prc}, providing valuable further information on the $^{25}$Al($p,\gamma$)$^{26}$Si entrance channel.

The radiative exit channel of the key 3$^+$ $^{25}$Al($p,\gamma$)$^{26}$Si resonance sets the resonance strength, but it has not yet been observed due to the dominance of the proton-decay channel, which is generally expected to be about 2 orders of magnitude stronger. Discovery of the exit channel via the strongest expected primary gamma-ray transition ($E_{\gamma} = 1739 \pm 2$ keV \cite{th04epj,au12cpc} or $1740 \pm 4$ keV \cite{ch10prc}) to the 3$^+$ level at 4187 keV \cite{se07prc} could lead to an experimental value for the small gamma-ray branching ratio $\Gamma_{\gamma}/\Gamma \approx \Gamma_{\gamma}/\Gamma_p$ of the 3$^+$ resonance. Together with $\Gamma_p$ \cite{pe09prc}, such a value would complete the experimental information on $^{26}$Si that is needed to calculate the resonance strength without relying heavily on shell-model calculations or properties of the mirror nucleus. Even a sufficiently strong upper limit could prove that the $^{25}$Al($p,\gamma$)$^{26}$Si reaction channel bypassing $^{26}$Al production is effectively closed in novae.

We have exploited the strong population of the third $3^+$ ($3^+_3$) $^{26}$Si level of interest in $^{26}$P beta decay \cite{th04epj,wr09prc} to search for the radiative exit channel and measure the gamma-ray branching ratio. Fast ions of $^{26}$P were produced at Michigan State University's National Superconducting Cyclotron Laboratory (NSCL) using projectile fragmentation of a 150 MeV/\textit{u}, 75 pnA $^{36}$Ar primary beam from the Coupled Cyclotron Facility, incident upon a 1.55 g/cm$^2$ Be transmission target. The $^{26}$P ions were separated from other fragmentation products by magnetic rigidity using the A1900 fragment separator \cite{mo03nim} (employing a 120 mg/cm$^2$ Al wedge) and by time of flight using a radio-frequency fragment separator \cite{ba09nim}. Up to 100 $^{26}$P ions s$^{-1}$ were delivered to the experimental setup. Clean ion identification was accomplished using both the time of flight from a scintillator at the focal plane of the A1900 to two 60 $\mu$m-thick silicon detectors located $\approx70$ cm upstream of the counting station, and the energy losses in the latter. On average, the  composition of the beam delivered to the experiment was found to be 74\% $^{26}$P with 18\% contamination by $^{24}$Al and small fractions of lighter ions. The $^{26}$P ions were implanted in a 1-cm thick planar germanium detector (GeDSSD) \cite{la13nim} that was divided electronically by 16 segmented strips of 5 mm pitch on the front side and 16 orthogonal ones on the back. The GeDSSD recorded the radioactive ion implantations and their subsequent beta decays using parallel low- and high-gain amplifications, respectively. The SeGA array of Ge detectors \cite{mu01nim} surrounded the GeDSSD in two coaxial 13-cm radius rings consisting of eight germanium detectors apiece and was used to detect gamma rays. The NSCL digital data acquisition \cite{st09nim} was employed.

The SeGA spectra were gain matched to produce cumulative spectra with 1-keV-wide bins using the strong gamma-ray lines from room-background activity at 1460.8 keV (from $^{40}$K decay) and 2614.5 keV (from $^{208}$Tl decay) as reference points, providing an \textit{in situ} first-order relative-energy calibration. Efficiency calibrations were performed using standard sources of $^{154,155}$Eu and $^{56}$Co placed along the beam axis on the outside surface of the GeDSSD cryostat (5 cm downstream of the $^{26}$P implantation position). The calibration data were used together with \textsc{geant4} Monte Carlo simulations incorporating the essential components of the experimental geometry to determine the efficiency at the $^{26}$P position 300 $\mu$m deep inside the GeDSSD crystal.

\begin{figure}
\includegraphics[width=0.5\textwidth]{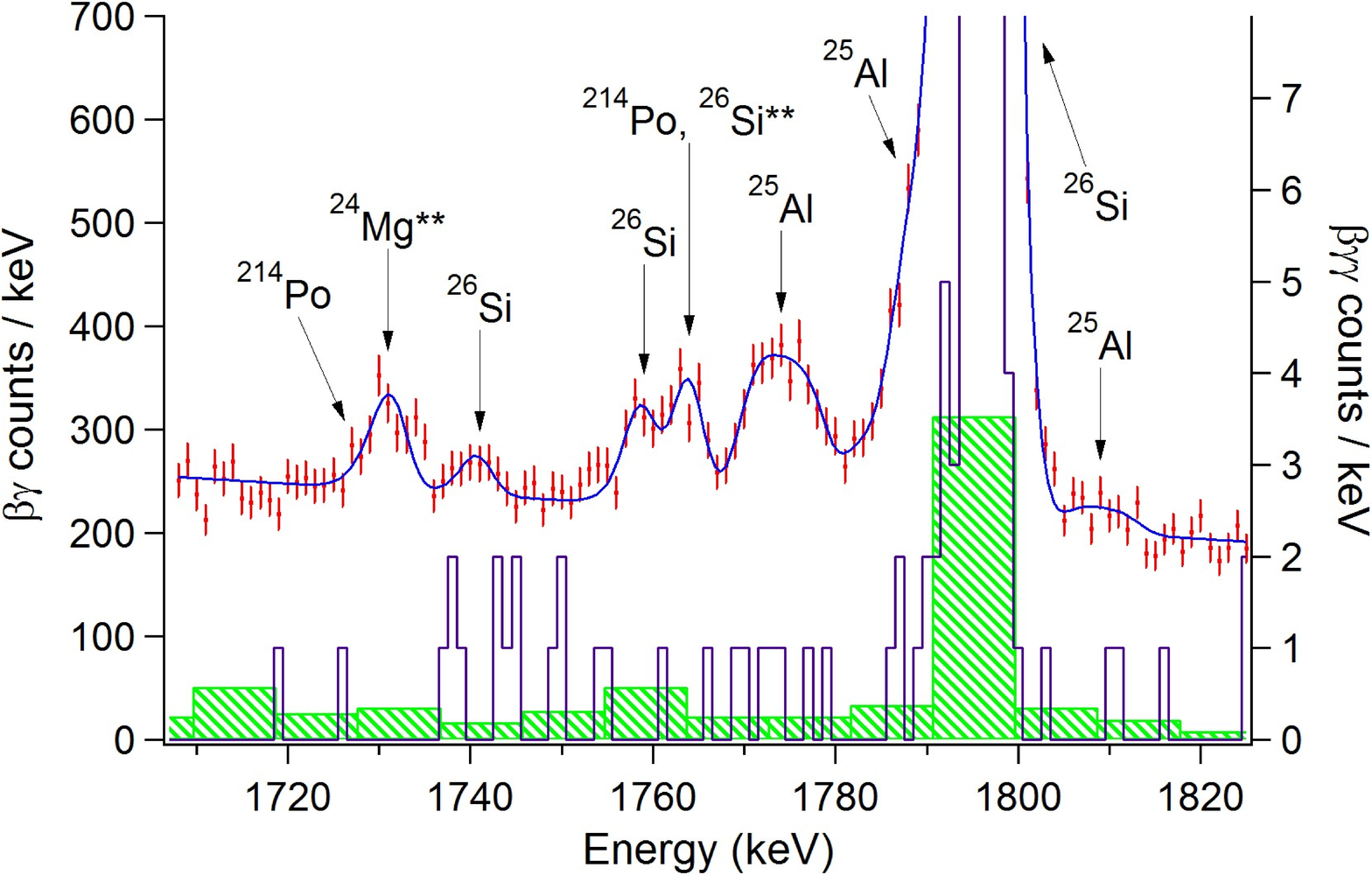}
\caption{(color online). Cumulative $^{26}$P($\beta\gamma$)$^{26}$Si (left axis) and $^{26}$P($\beta\gamma\gamma$)$^{26}$Si (right axis) spectra. The data points show the $\beta\gamma$ spectrum with error bars spanning 1 standard deviation. The solid line is a fit to the data including known gamma-ray transitions (Doppler broadened for $^{25}$Al), a straight-line background, and a new peak at 1742 keV. The gamma-ray emitting nuclides contributing to each feature in the spectrum are labeled, where two asterisks denote peaks produced by the escape of two 511-keV positron-annihilation gamma rays. The open histogram shows $\beta\gamma\gamma$ coincidences with 1401-keV gamma rays. The hatched histogram shows coincidences with continuum background in a relatively broad energy region just above 1401 keV in 9-keV-wide bins and normalized to correspond to the expected background per keV in the 1401-keV coincidence spectrum.
}
\label{fig: bigfit}
\end{figure}

In order to reduce the room-background contribution to the online gamma-ray spectra, a beta-delayed gamma-ray spectrum (see Fig. \ref{fig: bigfit}) was produced by requiring coincidences with high-gain events in the GeDSSD, which included beta particles, in a $1.2~\mu$s software gate. Although there were clear contributions from beam contaminants, room background, and daughter activities, this $\beta\gamma$ spectrum was dominated in the region of interest by $^{26}$Si and $^{25}$Al lines from the decay of $^{26}$P.

The spectrum was fit (see Fig. \ref{fig: bigfit}) in the region of interest using an exponentially modified Gaussian effective response function whose shape was fixed based on peaks of similar energy in the high statistics gamma-ray singles spectrum. A small extra peak was needed at 1801 keV to achieve a reasonable fit, but we considered this to be part of the intense 1797-keV line, which could not be fit adequately with our simplified response model. Doppler broadening of peaks originating from the $^{26}$P($\beta p \gamma$)$^{25}$Al decay sequence was incorporated to account for nuclear recoil. Where relative energies or intensities of lines were sufficiently well known, they were constrained in the fit. The continuum background from Compton scattering of higher-energy gamma rays was modeled with a straight line spanning the range shown in Fig. \ref{fig: bigfit}; there was no evidence for significant curvature.

We observed evidence for an unbroadened peak in the region of interest that was 3.9 standard deviations above the expected background level. The energy of this peak was found to be $1741.6 \pm 0.6 (\textrm{stat}) \pm 0.3 (\textrm{syst})$ keV with reference to the literature energies of well-known gamma-ray lines included in the fit and the relative-energy scale derived from the gain-matching procedure, where the uncertainties are statistical and systematic, respectively. The systematic uncertainty was assigned based on small variations in the result obtained using various calibration points. We could not find a previously observed gamma-ray transition to attribute this peak to, but its energy and (as shown below) intensity are consistent with expectations for the previously unobserved $3_3^+ \rightarrow 3_2^+$ transition to the 4187-keV state in $^{26}$Si. When additional narrow peaks were added to the fit in the vicinities of 1735 and 1754 keV (neither one included in the fit shown in Fig. \ref{fig: bigfit}), they were found to be 2.6 standard deviations above the expected background level. The 1742-keV peak is the most statistically significant new peak in the spectrum and it is consistent with the expected energy for the $3_3^+ \rightarrow 3_2^+$ transition.

In order to further test the hypothesis that the new 1742-keV peak is from the $3_3^+ \rightarrow 3_2^+$ $^{26}$Si gamma-ray transition following the beta decay of $^{26}$P, we searched for gamma-ray coincidences with the 1401-keV gamma ray, which is known to be the strongest transition deexciting the $3_2^+$ level at 4187 keV \cite{se07prc}. Supposing that the 1742-keV peak in the $\beta\gamma$ spectrum was from the $3_3^+$ level and using its measured intensity together with the known branching \cite{se07prc,wr09prc} and detection efficiency for the 1401-keV gamma ray, we would expect to observe $3.4 \pm 0.9(\textrm{stat}) \pm 0.4(\textrm{syst})$ real coincidences between 1401- and 1742-keV gamma rays. Figure \ref{fig: bigfit} shows the observed $\beta\gamma\gamma$ coincidence spectrum with nine candidate events over a 10-keV range that could be reasonably attributed to the 1742-keV peak in the $\beta\gamma$ spectrum. We estimate the background to be 0.3 counts/keV using both coincidences in nearby bins and the spectrum observed in coincidence with a relatively broad background region just above 1401 keV, suggesting an expected background of three counts over the 10-keV range. Assuming a Poisson distribution, the probability of obtaining nine or more counts when three background counts are expected is only 0.4 \%. This 99.6 \% confidence-level excess in the vicinity of the 1742-keV peak is the most statistically significant coincidence signal in the region of interest with the exception of the signal from the 1797-keV peak, for which the corresponding gamma ray is known \cite{th04epj,se07prc} to be emitted in cascade with the 1401-keV gamma ray. Considering the expected coincidence background of three counts and the nine observed events, we find the observed number of real coincidences to be $6.0^{+3.8}_{-2.7}$, where the uncertainties are adopted from the tables of Ref. \cite{fe98prd}. This $\beta\gamma\gamma$ coincidence result is consistent with the hypothesis that the 1742-keV peak in the $\beta\gamma$ spectrum is produced by deexcitation of the $3_3^+$ $^{26}$Si level, providing further evidence for such an identification. 

While no individual piece of evidence on the identification of the 1742-keV gamma ray (energy, intensity, coincidences) is absolutely conclusive on its own, consideration of the evidence as a whole presents a relatively strong case that this gamma ray is from the $3_3^+ \rightarrow 3_2^+$ $^{26}$Si transition. The most significant excess in the region of interest of both the $\beta\gamma$ and $\beta\gamma\gamma$ spectra is at 1742 keV, which is consistent with the expected transition energy. The intensities of the signals at 1742 keV in the two spectra are mutually consistent. The $\beta\gamma$ feature at 1754 keV and, to a lesser extent the one at 1735 keV, are inconsistent with the transition energy. In addition, these less statistically significant $\beta\gamma$ features do not have significant corresponding excesses in the $\beta\gamma\gamma$ spectrum. We will, therefore, consider the 1742-keV gamma ray to be from the $3_3^+ \rightarrow 3_2^+$ $^{26}$Si transition for subsequent calculations.

Using the fits of the 1742- and 1797-keV peaks in the $\beta\gamma$ spectrum shown in Fig. \ref{fig: bigfit} \cite{th04epj,se07prc}, the ratio of their respective areas was derived to be $(3.4 \pm 0.9) \times 10^{-3}$. Using this value, the known intensity of the 1797-keV line ($52\% \pm 11 \%$ \cite{th04epj}), and the ratio of efficiencies between these two energies (effectively unity) we derive the $\beta\gamma$ intensity of the 1742-keV gamma ray to be $[0.18 \pm 0.05(\textrm{stat}) \pm 0.04(\textrm{lit.})] \%$, where the latter uncertainty is due to adopted literature data. The 1742-keV partial gamma-decay branch of the $3_3^+$ level is expected to be $71^{+13}_{-19}\%$ based on the decay of the $^{26}$Mg mirror level \cite{wr09prc} (the shell model also predicts 71\% \cite{br13priv}), suggesting that the total $\beta\gamma$ intensity of all primary gamma rays from this level is $[0.25 \pm 0.07(\textrm{stat})^{+0.08}_{-0.07}(\textrm{lit.})]\%$. For comparison, the beta delayed proton decay intensity through this level is $17.96\% \pm 0.90 \%$ \cite{th04epj}. Dividing the gamma intensity by the proton intensity yields the ratio of partial widths, $\Gamma_{\gamma}/\Gamma_p= 0.014 \pm 0.004 (\textrm{stat})^{+0.005}_{-0.004}(\textrm{lit.})$. Adopting the experimentally determined value of $\Gamma_p = 2.9 \pm 1.0$ eV \cite{pe09prc} yields $\Gamma_{\gamma} = 40 \pm 11(\textrm{stat})^{+19}_{-18}(\textrm{lit.})$ meV allowing us to calculate a $^{25}$Al($p,\gamma$)$^{26}$Si resonance strength $\omega\gamma = 23 \pm 6 (\textrm{stat}) ^{+11}_{-10}(\textrm{lit.})$ meV.

The presently derived $^{25}$Al($p,\gamma$)$^{26}$Si resonance strength is the first one based on measurements of $^{26}$Si partial widths and branching ratios. The shell model has been used elsewhere \cite{ri11prc} to predict a resonance strength of 68 meV, whereas estimates \cite{ba02prc,wr09prc} based on a lifetime measurement \cite{gl86zpa} of the $^{26}$Mg mirror level yield a value of $18^{+18}_{-9}$ meV. The resonance strength derived in the present work favors those based on the mirror level.

We derive a $^{25}$Al($p,\gamma$)$^{26}$Si resonance energy for the $3_3^+$ level using our measured primary gamma-ray energy by adding it to the excitation energy of the $3_2^+$ state ($4187.1 \pm 0.3$ keV) \cite{se07prc}, which yields an excitation energy of $5928.7 \pm 0.6(\textrm{stat}) \pm 0.3 (\textrm{syst}) \pm 0.3 (\textrm{lit.})$ keV. Combining this with the $Q$ value yields a resonance energy of $414.9 \pm 0.6(\textrm{stat}) \pm 0.3 (\textrm{syst}) \pm 0.6(\textrm{lit.})$ keV. This energy is compatible with, and more precise than, the values of $412 \pm 2$ keV derived from the beta-delayed proton-decay energy \cite{th04epj,wr09prc} and $413 \pm 4$ keV from the ($p,t$) reaction \cite{ch10prc}.

We calculated a new thermonuclear $^{25}$Al($p,\gamma$)$^{26}$Si reaction rate using the $3^+$ resonance energy and strength (and corresponding uncertainties) from the present work. For the $1^+$ and $0^+$ resonances and the direct-capture component, we adopted the values and uncertainties from Ref. \cite{wr09prc}. We simulated the production of $^{26}$Al in novae on oxygen-neon (ONe) white dwarfs with masses of 1.15, 1.25, and 1.35 $M_{\odot}$ using the spherically symmetric, Lagrangian, hydrodynamic code \textsc{shiva} \cite{jo98apj} and a nuclear reaction network that includes the rates from Ref. \cite{il10npaII} and our new $^{25}$Al($p,\gamma$)$^{26}$Si rate. The simulations were repeated with the standard-deviation limits on our rate. The results are summarized in Table \ref{tab:26Al}, which shows that the uncertainties related to the $^{25}$Al($p,\gamma$)$^{26}$Si reaction are typically $\ll10\%$. Since the $^{25}$Al($p,\gamma$)$^{26}$Si reaction-rate uncertainty was the last substantial experimental nuclear-physics uncertainty associated directly with the explosion, the reported $^{26}$Al yields represent model predictions that are effectively independent of these experimental nuclear-physics uncertainties --- a significant milestone.  

Following the estimate of Ref. \cite{jo97apj} (based on Ref. \cite{we90aa}) and changing only the amount of $^{26}$Al produced in novae on 1.15 $M_{\odot}$ ONe white dwarfs (see Table \ref{tab:26Al}), we find an increase from 20\% to 30\% in the maximum contribution of classical novae to the $^{26}$Al observed \cite{di06nat,wa09aa} in the Milky Way. In order to deduce the nova contribution more accurately, the number of ONe novae per year in the Galaxy needs to be determined more accurately and multidimensional aspects of nova modeling, including mixing at the core-envelope interface, need to be integrated with the nucleosynthesis.

\begin{table}
\caption{\label{tab:26Al} Mass ejected from \textsc{shiva} \cite{jo98apj} simulations of novae occurring on oxygen-neon white dwarfs of various masses. $M_{\textrm{tot}}$ is the total mass ejected in a single outburst; $M$($^{26,27}$Al) are the masses of $^{26,27}$Al ejected. The uncertainties shown include only the effects of the standard deviation of the $^{25}$Al($p,\gamma$)$^{26}$Si reaction rate from the present work. The uncertainties in parentheses represent the results when the lower limit on the $^{25}$Al($p,\gamma$)$^{26}$Si reaction rate is calculated with the 3$^+$ resonance strength set equal to zero.}

\begin{ruledtabular}
\begin{tabular}{c c c c}
White-dwarf mass  & 1.15 $M_{\odot}$ & 1.25 $M_{\odot}$ & 1.35 $M_{\odot}$ \\
 \hline
 $M_{\textrm{tot}}$ ($10^{28}$ g)                & $4.9$                     & $3.8$                      & $0.90$                     \\
 $M$($^{27}$Al)/$M_{\textrm{tot}}$~($10^{-4}$)   & $85^{+1(+0)}_{-0}$        & $45^{+0(+0)}_{-0}$         & $32^{+1(+20)}_{-1}$        \\
 $M$($^{26}$Al)/$M_{\textrm{tot}}$~($10^{-4}$)   & $9.9^{+0.0(+0.0)}_{-0.1}$ & $5.8^{+0.0(+0.0)}_{-0.1}$  & $5.2^{+0.4(+4.8)}_{-0.3}$  \\
\end{tabular}
\end{ruledtabular}
\end{table}

In conclusion, we have observed evidence for a new $^{26}$P beta-delayed gamma ray at 1742 keV. The gamma-ray energy and intensity are consistent with those expected for the strongest primary transition deexciting the $3_3^+$ $^{26}$Si level. Coincidences with the secondary gamma ray at 1401 keV provide further evidence for such an identification. This is the first experimental evidence for the exit channel of the key $3^+$ resonance in the thermonuclear $^{25}$Al($p,\gamma$)$^{26}$Si reaction rate, which influences the production of $^{26}$Al in classical-nova models.  Using the energy and intensity of the observed $^{26}$Si gamma-ray line, we have derived the resonance energy and strength, allowing us to estimate $^{26}$Al production in novae in a manner that is effectively independent of nuclear-physics uncertainties. We checked the sensitivity of $^{26}$Al production to our lower limit by running our nova simulations with the 3$^+$ resonance strength set equal to zero and found that that only the simulation employing a $1.35 M_{\odot}$ white dwarf displayed any change (see Table \ref{tab:26Al}). Therefore, one could also interpret our experimental result as an upper limit and reach essentially the same astrophysical conclusions: our experiment is sufficiently sensitive to prove for the first time that the $^{25}$Al($p,\gamma$)$^{26}$Si reaction rate is very slow so that the path bypassing $^{26}$Al production is closed in novae hosted by typical oxygen-neon white dwarfs with masses below $1.3 M_{\odot}$ and only open for a short period of time near peak temperature for higher white-dwarf masses, which are expected to be scarce according to stellar evolution models.


We encourage future measurements to further reduce the uncertainties in the 3$^+$ $^{25}$Al($p,\gamma$)$^{26}$Si rate, which are dominated by the resonance strength uncertainty. For example, a $5\sigma$ detection of the 1742-keV gamma ray would be an improvement and higher-statistics data on $\gamma\gamma$ coincidences are desirable. First evidence for the weaker primary gamma-decay branches from the 5929-keV $^{26}$Si level could provide direct experimental constraints on the $3_3^+ \rightarrow 3_2^+$ branch we adopted from Ref. \cite{wr09prc}. Independent measurements of $\Gamma_p$ could be conducted to confirm the existing value \cite{pe09prc} and improve the uncertainty. Direct measurements of the $3^+$ resonance with intense low-energy $^{25}$Al beams will hopefully be feasible at the next generation of rare-isotope facilities; the present results provide essential information for the planning of such experiments.

This work was supported by the U.S. National Science Foundation under Grants No. PHY-1102511 and No. PHY 08-22648, the U.S. Department of Energy under Contract No. DE-FG02-97ER41020, the U.S. National Nuclear Security Agency under Contract No. DE-NA0000979, MEC Grant No. AYA2010-15685, and the ESF EUROCORES Program EuroGENESIS through Grant No. EUI2009-04167. We gratefully acknowledge A. Garc\'{\i}a for advice during the preparation of our experimental proposal and the NSCL Operations staff for delivering the beam.

\end{document}